\documentclass[amsmath,amssymb,aps,prd,showpacs,nofootinbib]{revtex4}

\usepackage{graphicx}
\begin{document}
\title{Proton and neutron form factors with quark orbital excitations}

 \author{\firstname{Yu.A.}~\surname{Simonov}}
\email{simonov@itep.ru} \affiliation{NRC ``Kurchatov Institute'' -- ITEP, B. Cheremushkinskaya 25, Moscow, 117259, Russia}







\newcommand{\beq}{\begin{eqnarray}}
 \newcommand{\eeq}{\end{eqnarray}}
\newcommand{\be}{\begin{equation}}
 \newcommand{\ee}{\end{equation}}

 \def\la{\mathrel{\mathpalette\fun <}}
\def\ga{\mathrel{\mathpalette\fun >}}
\def\fun#1#2{\lower3.6pt\vbox{\baselineskip0pt\lineskip.9pt
\ialign{$\mathsurround=0pt#1\hfil ##\hfil$\crcr#2\crcr\sim\crcr}}}
\newcommand{\veX}{\mbox{\boldmath${\rm X}$}}
\newcommand{{\SD}}{\rm SD}
\newcommand{\pp}{\prime\prime}
\newcommand{{\Mc}}{\mathcal{M}}
\newcommand{\veY}{\mbox{\boldmath${\rm Y}$}}
\newcommand{\vex}{\mbox{\boldmath${\rm x}$}}
\newcommand{\vey}{\mbox{\boldmath${\rm y}$}}
\newcommand{\ver}{\mbox{\boldmath${\rm r}$}}
\newcommand{\vesig}{\mbox{\boldmath${\rm \sigma}$}}
\newcommand{\vedelta}{\mbox{\boldmath${\rm \delta}$}}
\newcommand{\veP}{\mbox{\boldmath${\rm P}$}}
\newcommand{\veA}{\mbox{\boldmath${\rm A}$}}
\newcommand{\vep}{\mbox{\boldmath${\rm p}$}}
\newcommand{\veq}{\mbox{\boldmath${\rm q}$}}
\newcommand{\veQ}{\mbox{\boldmath${\rm Q}$}}
\newcommand{\vez}{\mbox{\boldmath${\rm z}$}}
\newcommand{\veS}{\mbox{\boldmath${\rm S}$}}
\newcommand{\veL}{\mbox{\boldmath${\rm L}$}}
\newcommand{\vel}{\mbox{\boldmath${\rm l}$}}
\newcommand{\veR}{\mbox{\boldmath${\rm R}$}}
\newcommand{\ves}{\mbox{\boldmath${\rm s}$}}
\newcommand{\vek}{\mbox{\boldmath${\rm k}$}}
\newcommand{\ven}{\mbox{\boldmath${\rm n}$}}
\newcommand{\veu}{\mbox{\boldmath${\rm u}$}}
\newcommand{\vev}{\mbox{\boldmath${\rm v}$}}
\newcommand{\veh}{\mbox{\boldmath${\rm h}$}}
\newcommand{\vew}{\mbox{\boldmath${\rm w}$}}
\newcommand{\verho}{\mbox{\boldmath${\rm \rho}$}}
\newcommand{\vexi}{\mbox{\boldmath${\rm \xi}$}}
\newcommand{\veta}{\mbox{\boldmath${\rm \eta}$}}
\newcommand{\veB}{\mbox{\boldmath${\rm B}$}}
\newcommand{\veH}{\mbox{\boldmath${\rm H}$}}
\newcommand{\veE}{\mbox{\boldmath${\rm E}$}}
\newcommand{\veJ}{\mbox{\boldmath${\rm J}$}}
\newcommand{\veal}{\mbox{\boldmath${\rm \alpha}$}}
\newcommand{\vepi}{\mbox{\boldmath${\rm \pi}$}}
\newcommand{\vegam}{\mbox{\boldmath${\rm \gamma}$}}
\newcommand{\vepar}{\mbox{\boldmath${\rm \partial}$}}
\newcommand{\llan}{\langle\langle}
\newcommand{\rran}{\rangle\rangle}
\newcommand{\lan}{\langle}
\newcommand{\ran}{\rangle}


\Large

\begin{abstract}
Nucleon form factors play an especially important role in studying the  dynamics of nucleons
and explicit structure of the wave functions at arbitrary nucleon velocity. The purpose of the paper is to explain theoretically all four nucleon form factors  measured experimentally  in the cross section measurements (by the Rosenbluth method), yielding almost equal normalized form factors $G^p_E,G^p_M,G^n_M$), as well as in the polarization transfer experiments, where a strongly decreasing proton electric form factor has been discovered. It is shown, using relativistic hyperspherical formalism, that the nucleon wave functions in the lowest (hypercentral) approximation provide almost equal normalized form factors as seen in the Rosenbluth cross sections, but in the higher components they contain a large admixture of the quark orbital momenta, which strongly decreases $G^p_E$ and this effect is possibly detected in the polarization transfer method (not seen in the classical cross section experiments). Moreover, the same admixture of the higher components explains the small positive form factor $G^n_E$, which is zero in the hypercentral approximation.
 The resulting  form factors, $G^p_M(Q),G^p_E(Q),G^n_M(Q)$ are calculated up to $Q^2\approx 10$~GeV$^2$, using the  the Lorentz contracted nucleon wave functions  and shown to be  in reasonable agreement with experimental data.

\end{abstract}
      \maketitle

 \section{Introduction}

The baryon form factors have been a specific source of knowledge of the internal hadron structure for a long time. Since the Hofstadter experiments \cite{1,2} this field becomes a privileged meeting point where the theory can check its arguments vs  the experimental data, which is now reflected in the well- known publications \cite{3,4,5} and famous books \cite{6,7}. On the experimental side one has numerous measurements of baryon form factors (ff), which are reviewed and discussed in \cite{8,9} for the time-like and the space-like cases and in \cite{10} for the space-like case. An extensive analysis and reviews of the baryon form factors are given in \cite{11,12}. These reviews and  \cite{8} also contain  detailed discussion of the theoretical approaches to the problem of the baryon ff, which shall be used in what follows. One should note the growing role of the lattice measurements for the baryon ff,see e.g \cite{12*,13*,14*,15*}. The experimental results, obtained before the end of the previous century, were  using the cross section data for the form factors (the so-called Rosenbluth method) which have yielded three  almost equal normalized form factors for proton and neutron, except for a tiny electric neutron form factor, which could be readily explained in different theoretical models (see e.g. discussion in \cite{7,8,9,10}). Moreover, this equality is in line with the perturbative QCD predictions for the almost constant ratio $R(Q)= \frac{\mu_p G^p_E}{G^p_M}$  at larger $Q$ \cite{13}.

However, in a new series of experiments, using both the polarized electron  beam and target \cite{14,15},  the ratio $R(Q)$ was found to decrease fast with a possible zero value around $Q^2= 8$ GeV$^2$. This result was carefully checked and the detailed discussion of this phenomenon was given in \cite{16}. The possible theoretical mechanisms for this phenomenon were presented in \cite{17,17*} and one of the suggested reasons was the role of the internal quark orbital momenta in the nucleon, which possibly is not properly evident in the cross section experiments. It is the purpose of the present paper to develop the formalism which can account for nonzero admixture of internal quark orbital momenta and predict the resulting behavior of all four nucleon form factors - $G^p_E, G^p_M, G^n_E, G^n_M$ as the functions of $Q$.

In what follows we shall be using  for baryon wave functions the instantaneous  formalism, based on the
results obtained with the relativistic QCD Hamiltonian, namely

1) a baryon can be described at all $Q$ by a small set of the wave functions, which are the eigenfunctions of the same QCD Hamiltonian with instantaneous interaction, obtained from relativistic path integral \cite{18}. This method was extensively used to calculate the meson masses and wave functions \cite{19,20,21} and has allowed to obtain numerous data with good agreement with experiment \cite{22,23}.

2) For baryons the dynamics is different from the two-body meson case, since the basic   interaction is
the instantaneous three-body string-like, and the  most reliable method, taking into account the three -body type of strong interaction in nucleon, is the hyperspherical method (HM). The HM for fermions was suggested as a first version in \cite{24'} and then developed in 1966 in the framework of non-relativistic nuclear physics and successfully applied for the nuclei  with three or more nucleons  \cite{24,25,26}. In the case of baryons one needs the relativistic version of the HM and it was developed in \cite{27,28,29,30,31,32}. In particular, in \cite{28}  for the first time the baryon Regge trajectory was obtained. Also the magnetic moments of baryons were found in \cite{29} with accuracy of 10\%.

Meanwhile the hypercentral version of the HM was actively developed in a large series of papers, first in the nonrelativistic hypercentral constituent quark model (NHCQM) (see \cite{33} and references therein) and later in the relativised form (RHCQM) \cite{34,35}. One should note that both, in the nonrelativistic hypercentral HM and in the RHCQM, one obtains naturally similar forms  $G^p_E(Q)$  and $\frac{G^p_M(Q)}{\mu_p}$,
but their ratio is difficult to make decreasing with $Q^2$. As we will show below, this fact is connected to the absence of internal quark orbital momenta in the hypercentral approximation and one needs to take into account the non-hypercentral contribution. One of the ways is to exploit the quark-diquark model, as it was done in \cite{35}.

In our case we shall use -- what  possibly  is the simplest way -- to extend the hyperspherical basis in our calculations. As it will be shown, it is enough to add to the basic hypercentral component the lowest noncentral contribution with the proper relative amplitude, given by the constant $A$, and it recovers both reasonable form factor $G^n_E$ and the ratio $R(Q)$. It is important that the sign and value of $A$ can be obtained also from  the HM formalism.

To ensure the correct $Q$ dependence of all four form factors, including large $Q$ values, one can use any of two strategies: 1. to find the proper set of parameters, defining the nucleon wave function and then to include the perturbative formalism \cite{13} to modify expressions at high $Q$;  2. to try as an alternative the Poincare` invariant instantaneous formalism with nonperturbative Lorentz contracted wave functions. The latter  are not destroyed by the high $Q$ photon since the highly accelerated (velocity $v$) proton wave functions have stabilized momentum dependence at high $Q$- $\psi_v(Q)=\psi_0(Q\sqrt{1-v^2})$ and can exceed the perturbative values. Therefore the accelerated Lorentz contracted wave function can escape the perturbative dissociation into unbound pieces as it is assumed in the perturbative approach \cite{13}
 Below we are using the formalism (2), based on the analysis of the relativistic dynamics  in QCD and QED with the notion of the Lorentz contracted wave functions, introduced  in \cite{36,37,38}, where it was checked for real QCD systems, and  exploited in \cite{39} for the meson form factors, showing  a good agreement with experiment, and in \cite{40} for the amplitudes of the strong decays, explaining the old problem in $\rho,\rho'$ decays. It is essential that in this formalism one can use the same set of the wave functions in the rest frame for all values of Q, taking into account the transformation laws of the Lorentz contracted wave functions.
One should note at this point that the approach using relativistically transformed wave functions for the baryon
form factors was possibly exploited for the first time by Licht and Pagnamenta \cite{41} in 1970, as will be discussed in the next section. We are using below a modified form, suitable for 3 and more constituents in the wave function.
We shall show that our expressions for all four form factors, containing only two parameters (numbers), agree reasonably well with available experimental data, and what is remarkable, the Rosenbluth type data agree with our results for the zero orbital admixture $A$, while the polarization transfer data agree with the fixed nonzero value of $A$. In addition to this surprising result we shall demonstrate  a remarkable structure of the Lorentz contracted  nucleon form factor, which allows to continue its form up to large $Q$, extending in this way the positive experience with the meson form factors \cite{39} and the meson decay amplitudes \cite{40}.

The paper is organized as follows. In the next section we outline the general structure  and dynamical variables of the baryon wave functions in the hyperspherical formalism. In section 3 the electromagnetic ff of both proton and neutron are defined via the wave functions and expressed via the only function in the first approximation, when the dominant symmetric part of the coordinate and the spin-flavor functions are retained.  In section 4 a simple Gaussian form with the only parameter for all space-like ff is compared with experimental data in the whole available interval of $Q^2$. In section 5 the hyperspherical formalism is exploited to find the nucleon wave functions including the non-symmetric part of the wave functions, providing in this way the neutron electric $G_E^{n}$ ff. In the section 6 the comparison of the resulting non-symmetric form factors  with experimental data is given. The last section contains the conclusions and outlook. The Appendix A1  contains additional material on the form factors and numerical data on matrix elements.

\section{General structure of the baryon  wave functions}
Below, as well as in the basic papers \cite{36,37,38}, we adopt  a very simple idea that
1) a baryon can be described at all momenta $Q$ by a small set of the wave functions, which are the eigenfunctions of the same QCD Hamiltonian with instantaneous interaction, which was obtained from the relativistic path integral \cite{41*,42,43,21};
2) the same set of the wave functions can be used for all values of $Q$, taking into account the Poincare` transformation laws of the instant form wave functions, yielding Lorentz contracted form of these functions in addition to the proper factors for spinors etc. This main point of the present paper  is based on the analysis of relativistic dynamics in an arbitrarily moving system done in \cite{36}. The role of the Lorentz contraction of the wave functions for the meson ff was studied in \cite{39}, where it was shown that the resulting ff are obtained to be in good agreement with experiment. Moreover, the same Lorentz contraction effect explains the unusual $\Gamma(s)$ dependence
of the $\rho,\rho'$ meson decays \cite {40}. Also  the account of the Lorentz contraction in the pdf allows to solve the old problem of the proton spin \cite{38}.

The main point of this approach is the fundamental invariance law of the probability density $\rho(\vex,t)$ in the volume element $dV$ for a velocity $\vev$ \cite{44}, \cite{45}
\be
\rho(\vex,t)dV = {\rm invariant},~~\label{eq.1}
\ee
where $\rho(\vex,t)$ is the density, associated with the wave function $\psi_n(\vex,t)$. Here $\vex$ is the chosen coordinate, characterizing the density distribution in a hadron; for a many-particle system all other interparticle d.o.f are assumed to be integrated out, and the density can be written as
\be
\rho_n(\vex,t) = \frac{1}{2i} \left(\psi_n \frac{\partial \psi_n^+}{\partial t}  - \psi_n^+ \frac{\partial \psi_n}{\partial t}\right)
= E_n |\psi_n(\vex,t)|^2, ~~\label{eq.2}
\ee
and $dV=d\vex_{\bot} dx_{\|}$. One can use the standard transformations,
\be
L_{\rm P}dx_{\|} \rightarrow dx_{\|} \sqrt{1 - \vev^2}, ~~ L_{\rm P} E_n \rightarrow \frac{E_n}{\sqrt{1-\vev^2}},
\label{eq.3}
\ee
to insure the invariance of  (\ref{eq.1}). In its turn the invariance law implies that in the wave function $\psi(\vex,t)=\exp(-iE_nt)\varphi_n(\vex)$ the function $\varphi_n(\vex)$ is deformed in the moving system,
\be
L_{\rm P}\varphi_n(\vex_\bot, x_{\|}) = \varphi_n\left(\vex_\bot, \frac{x_{\|}}{\sqrt{1-\vev^2}}\right),
\label{eq.4}
\ee
and can be normalized as
\be
\int E_n |\varphi_n^{(v)}(\vex)|^2 dV_v = 1 = \int M_0^{(0)} |\varphi_n^{(0)}(\vex)|^2 dV_0,
\label{eq.5}
\ee
where the subscripts $(v)$ and $(0)$ refer to the moving and the rest frames. One of the immediate consequences from the
Eqs.~(\ref{eq.3}) and (\ref{eq.4}) is the property of the boosted Fourier component of the wave function:
\be
\varphi_n^{(v)}(\veq) = \int \varphi_n^{(v)}(\ver) \exp(i\veq\ver) d\ver = C_0 \varphi_n^{(0)}(\veq_\bot, q_{\|}\sqrt{1-v^2}),
\label{eq.6}
\ee
where $C_0 = \sqrt{1-v^2} = \frac{M_0}{\sqrt{M_0^2 + \veP^2}}.$

 We have considered above the case of one coordinate vector in the wave function, where one can associate $\vex$ for a meson wave function with the relative coordinate of quark and antiquark, while for the baryon one has two relative coordinates $\vexi,\veta$, which are transformed in the same way. To take into account all participating coordinates, as it appears in the multipartice correlation functions in nuclei,
 one must consider Lorentz contraction of the wave function with respect to all of them, as it was done in the pioneering paper of Licht and Pagnamenta in 1970 \cite{41}. However for the form factor, where  photon interacts
 with only one quark at a time, we must  take into account in (\ref{eq.5}) the one-coordinate density and integrate out all other coordinates in a Lorentz invariant form. One of possible ways is the account of all extra
 longitudinal coordinates via a scalar Lorentz invariant ratio : $ a_i= \frac{\xi_{\|}^{(i)}}{\xi_{\|}^{(1)}}$,
 with $i= 2,3,..$. In this way the baryon form factor obtains the same power of $C_0$ as the meson form factor, and
 we  exploit this approach in the paper. This is in contrast to the form used in \cite{41}.
\be
\varphi_n^{(v)}(\veq,\vek) = \int \varphi_n^{(v)}(\vexi,\veta) \exp(i ({\veq\vexi + \vek\veta}))d\vexi d\veta=
C_0 \varphi_n^{(0)}(\veq_\bot,q_{\|}\sqrt{1-v^2}, \vek_\bot,k_{\|}\sqrt{1-v^2}),
\label{eq.7}
\ee
It is clear that these wave functions in the Breit frame enter the baryon ff in the following way,
\be
F_a(Q) = C_0 \sum_i \frac{e_i}{e} \int d^3 q d^3 k \varphi_{(-Q/2)}(\veq,\vek) \varphi_{(Q/2)}(\veq +\nu_1
\veQ, \vek +
\nu_2 \veQ)
\label{eq.8}
\ee
Here $\nu_i$ are chosen for the parts of $\veQ$ absorbed by quarks. Using now (\ref{eq.7}) one can see that the momentum $\veQ$ in the quark wave function is multiplied by $\sqrt{1-v^2} = M/\sqrt{M^2 + Q^{2}/4}$ and is always of the order of M. As a result, one never enters in the high $Q$  area, where one expects the perturbative asymptotics. In addition to this surprising result above it was demonstrated  a remarkable structure of the nucleon ff,
which allows  in the first approximation to consider the unique standard form, valid for all expressions.

To define explicitly the variables and interaction in our formalism, one can use the baryon Hamiltonian as it was formulated in \cite{27,28,29,30,31} together with a spin-dependent interaction \cite{30}, which is an extension of the method, based on the path integral Fock--Feynman--Schwinger Representation (FFSR) \cite{41*}. For the mesons the corresponding Hamiltonian was obtained in \cite{42} and with inclusion of external electromagnetic field in \cite{43}. The numerous meson spectra, obtained within this method, have been discussed in \cite{27,20,21} and recently reviewed in \cite{22,23}. We write the total Hamiltonian of the baryon as
\be
 H_{RQM} = \sum_i \left( \sqrt{p_i^2 + m_i^2} + \sigma \left|\vec r_i\right|\right ) + V_{int}
\label{eq.9}
\ee
Here $m_i$ are true quark masses and one can introduce the quark energy $\omega_i$ in the c.m. of the baryon $\omega_i= \sqrt{p_i^2 + m_i^2}$ and write the total energy in terms of $\omega$ and momenta, as it was done in \cite{27,28,29}.
Then the  baryon spectra and wave functions can be obtained using the hyperspherical (or the K-harmonics) method \cite{24,25,26}, which was shown to be useful  in nuclear physics, but especially convenient in its relativistic version \cite{27,28,29,30,31} for the baryons, where the confinement interaction is of the tree-body character, and it allows to obtain not only the masses but also the baryon Regge trajectories \cite{28}. This approach and its results for the nucleon ff in the noncentral case will be the basic point of the detailed study of nucleon ff planned for the future. In the present paper we start with a simplified analysis using a parametrized form of baryon wave functions to answer two basic questions:\\
1) What is the basic effect of Lorentz contraction on the baryon ff?\\
2) What is the property of the baryon wave function, which can solve the discrepancy between the Rosenbluth and polarization transfer experiments?

In this section we shall start with the simplified (the hypercentral) version of the baryon dynamics and the wave functions to define its basic structure and static properties (the baryon mass, spin and magnetic moments). To simplify matter we shall put the quark masses $m_i=0$ and neglect the difference between the quark average energies, setting $\omega_i = \omega$. In this case the definitions of the relative coordinates and relative momenta in a baryon are following,
\be
\veR = 1/3 \sum_i \vez_i, ~\veta = \frac{ \vez^{(2)}-\vez^{(1)}}{\sqrt{2}},~~ \vexi = \frac{\vez^{(1)}+\vez^{(2)}-2\vez^{(3)}}{\sqrt{6}}
\label{eq.10}
\ee

Correspondingly, we define the momenta $\veq =  \frac{\partial}{i \partial \vexi}$  and
$\vek = \frac{\partial}{i \partial \veta}$, so that the kinetic part of the Hamiltonian can be written as
$H_{kin} = \frac{3 \omega}{2} + \frac{(\veq^2 + \vek^2)}{2\omega}$ and the confining part of interaction as
$V_{(\rm conf)} = \sigma \sum_i |\vez^{(i)} - \vez^{(Y)}|$, where $\vez^{(Y)}$ is the Torricelli point, which we shall take in the lowest approximation to coincide with the c.m. coordinate $\veR$. One can find explicit calculations of the nucleon mass in this formalism in \cite{31,32}, which yields the average mass of $N,\Delta$  equal  1.08 GeV, in good agreement with data.

We now come to the structure of the baryon wave function, which is crucial for the ff. We shall neglect in the first approximation the spin dependent interactions and consider the wave functions in a moving system depending on spin projection $\sigma$, quark flavor $\tau$ and quark coordinates $x_i$ or momenta $p_i$. The total baryon wave function, depending on the coordinates (x), spin-flavor $(\sigma, \tau)$ and the color, has the well-known structure in nonrelativistic case (and can be also applied to the upper parts of the bispinor  wave functions) \cite{48}
\be
\Psi_{\rm tot} = ( \varphi_{\rm sym}(x) \psi_{\rm sym}(\sigma, \tau) + (\varphi'(x) \psi'(\sigma,\tau) + \varphi''(x) \psi''(\sigma,\tau)) +
\varphi_a(x) \psi_a(\sigma,\tau) ) \chi_a{\rm (color)},
\label{eq.11}
\ee
where $\chi_a $ is the antisymmetric color function, $f_{\rm sym}$ and $f' ,f''$ are symmetric and mixed symmetry functions, while $\sigma, \tau$ refer to spin and isospin parts of wave functions. Here  is an important point: as the first approximation we shall keep only the symmetric coordinate wave function (and correspondingly, only symmetric spin-isospin part) and then find a strong simplification of all results for the baryon dynamics and especially for the form factors. Earlier in \cite{29} it was shown that the purely symmetric part of wave functions yields nucleon magnetic moments with a good accuracy, meaning that  neglected part contributes less than 10\% to the magnetic moments, while it becomes  very important for the form factors of nucleons. In particular, we shall show that the most important role of the mixed symmetry wave function
is that it ensures nonzero contribution to the neutron charge form factor and the difference between the proton charge form factor and the proton magnetic form factor divided by its magnetic moment, which is especially important for the explanation of the modern polarization-based experiments (see discussion below in section 6). At this point it is important to classify further the coordinate wave function and its possible symmetric and nonsymmetric parts with the help of the hyperspherical formalism \cite{24}, where the wave function can be written as
\be
\varphi(\vexi,\veta) = \frac{1}{ \rho^{2}} \sum_{K,\nu} u_K^{\nu}(\Omega) \psi_K^{(\nu)}(\rho).
\label{eq.12}
\ee
Here $\rho$ is the collective 3-quark coordinate, $\rho^2 = \sum_i (\vez^{(i)} - \veR)^2 = \vexi^2 + \veta^2$,
and $\Omega$ is a set of the angular coordinates, while for $K=0$ the angular function $u_0^{(0)}$ is a constant, implying that in the first term with $K=0$ in (\ref{eq.12}) -- in the hypercentral approximation -- it does not contain any angular momenta (since $\rho$ also has all angular momenta equal to zero. Hence, the first term is also fully symmetric in the quark coordinates, while the next term with $K=2$  already contributes to the mixed symmetry terms in  (\ref{eq.12}). Note also that the representation (\ref{eq.12}) can be written for any total orbital momentum $L$. In what follows we shall consider only $L=0$. In this case  all values $K$ are proportional to 2 and the term with $K=2$ contains both $\vexi\veta$ and $\veta^2 - \vexi^2$ terms with the orbital momenta $l_{\xi}=l_{\eta}=1$, with the zero sum: $\veL= \vel_\xi + \vel_\eta = 0$. One can estimate the relative contribution of the first terms. As it follows from the analysis in \cite{31}, already the $K=0$ term yields the mass of the nucleon with accuracy better than 5\%. Note also that the $K=0$ harmonic is concentrated
at small $\rho$ and should be more important with the growing $Q$.  In what follows we shall exploit this feature to find the nucleon ff, first in the $K=0$ approximation and then calculating the next terms. Another important check of our formalism is the calculation of the proton and neutron magnetic moments, done in the framework of our formalism in \cite{29}. Here the nucleon wave functions have been used in the simplest form.
Writing  the nucleon magnetic moment as $\mu_B = \lan\Psi_B|\mu_z|\Psi_B\ran$
and taking the baryon wave function in the  form
\be
\Psi_B = \varphi^{\rm(sym)}(z) \psi^{\rm(sym)}(\sigma,\tau) \chi({\rm color})
\label{eq.13}
\ee
and using the spin-flavor baryon wave functions   given in the Appendix A1,
 for $\sigma = 0.15$ GeV$^2 $  one obtains in \cite{29} the following proton and neutron magnetic moments
 \be
 ~~\mu_p = 2.54~ \mu_N , ~~\mu_n = - 1.69~ \mu_N,
 \label{eq.14}
 \ee
 which are by 10\% less than their experimental values and here will be considered as a  first measure of the accuracy of our approach for the nucleon ff.

 \section{The proton and neutron form factors in the symmetrical approximation}

In this section we consider the space-like nucleon ff  in the Breit frame, where the nucleon currents can be written as
\be
J^{(0)} = e 2 M \psi^{(+)} \psi ( F_1 - \tau F_2) = e 2M \psi^{(+)} \psi G_E^{N}
\label{eq.15}
\ee

\be
\veJ = ie \psi^{(+)} (\vesig \times \veQ) \psi (F_1 + F_2) = ie \psi^{(+)} (\vesig \times \veQ) \psi G_M^{N}.
\label{eq.16}
\ee

Then defining the symmetric coordinate wave function boosted with momentum $\veP$ as    $\psi_{\veP} (\vexi,\veta)$,
one can write in the Breit frame
\be
G^N_E = \sum_i \frac{e_i}{e} \int( d^3 \xi d^3 \eta \varphi_{-(Q/2)}(\vexi,\veta) \exp(i\veQ \ver_i) \varphi_{(Q/2)}(\vexi,\veta)).
\label{eq.17}
\ee
Now for the boosted wave function, as it is shown in the section 2 (\ref{eq.4})-(\ref{eq.8}), one has the relation
\be
\psi_{(\veQ/2)} ( \vexi,\veta) = \psi^{(0)} \left(\vexi_\bot,\xi_{\|}/\sqrt{1-v^2},~\veta_\bot,\eta_{\|}/\sqrt{1-v^2}\right).
\label{eq.18}
\ee
It is more convenient to write the ff in the momentum space, where we obtain, as in (\ref{eq.4})-(\ref{eq.8}), for the  $G^n_E(Q)$,
\be
G^n_E(Q) = C_0 \sum_i \frac{e_i}{e} \int\left( \frac{d^3 q d^3 k}{(2\pi)^6} \varphi_{(-Q/2)}(\veq,\vek) \varphi_{(Q/2)}(\veq +\nu_1 \veQ, \vek + \nu_2 \veQ)\right).
\label{eq.19}
\ee
Here $\nu_1^{(i)}, \nu_2^{(i)}$ define which part of $Q$ is brought by $q$ and $k$, respectively. From their definitions via $p_i$ one has
$\nu_1^{(i)} = 1/\sqrt6, 1/\sqrt6, -\sqrt{2/3} $ for $i = 1,2,3,$ and
$\nu_2^{(i)} = -1/\sqrt2, 1/\sqrt2, 0,$ for $i = 1,2,3.$
At this point one can formulate an important property of the hypercentral equations for the nucleon form factors.
Indeed, as it seen in (\ref{eq.17}),(\ref{eq.18}) the form factors are the Fourier transforms of the $K=0$ product of the
wave functions depending on $\rho=\sqrt{\vexi^2 + \veta^2}$ and therefore can depend only on the square of the six
dimensional vector $\nu_1^{(i)}\veQ,\nu_2^{(i)}\veQ$, i.e. on $[(\nu_1^{(i)})^2 + (\nu_2^{(i)})^2] \veQ^2$.
Note, that this property does not depend on the boost transformations and is valid both in the boosted and in the rest frame. Now using the explicit values of $\nu_1^{(i)},\nu_2{(i)}$ given above, this latter expression is equal to $\frac{2}{3}\veQ^2$ and does not depend on $i$. This allows to obtain below remarkable relations (\ref{eq.22}).

Now writing the boost transformation for the $\varphi_{(+/-Q/2)}$, one finally obtains for $G_E^p(Q)$
\be
G^p_E (Q) = C_0 \sum_i \frac{e_i}{e} \int \left(\frac{d^2 q d\chi d^2 k d\kappa}{(2\pi)^6} \varphi(\veq_\bot,\vek_\bot,\chi,\kappa)
\varphi(\veq_\bot,\vek_\bot,\chi + \nu_1^{(i)} Q C_0,\kappa + \nu_2^{(i)} Q C_0)\right).
\label{eq.20}
\ee
Here we have introduced the new integration variables $\chi = q_{\|} C_0$ and $\kappa = k_{\|} C_0$.
Since $C_0 = M/\sqrt{M^2 + Q^2 /4}$, one can see that $G_E^p (Q=0) = 1.0$.
At the same time one can see that the neutron ff vanishes $G_E^n(Q) = 0 $ for all Q values in this symmetric approximation, since $G_E^n(Q)= \sum_i \frac{e_i}{e} f(Q)$, and $f(Q)$ does not depend on $i$.
This is probably a natural explanation of the smallness of the neutron electric ff.
We now turn to the normalized magnetic ff's, $\frac{G_M^p}{\mu_p} $ and $\frac{G_M^n}{\mu_n}$, and shall show that they are equal and, moreover, they are also equal to $G_E^p$ in this symmetric approximation.
Indeed, one can write the nucleon magnetic ff as follows,
$$
G^N_M(Q) = \frac{M}{\omega} \sum_i \int \frac{d^2 q d\chi d^2 k d\kappa}{(2\pi)^6} \varphi(\veq_\bot,\vek_\bot,\chi,\kappa)\times$$\be \times\psi^{\rm(sym)}(\sigma,\tau) \frac{e_i \sigma_z^{(i)}}{e}
\psi^{\rm(sym)}(\sigma,\tau) \varphi(\veq_\bot,\vek_\bot,\chi + \nu_1^{(i)} Q C_0,\kappa + \nu_2^{(i)} Q C_0).
\label{eq.21}
\ee
Now using the the structure of the $\psi(\sigma,\tau)$, given in the Appendix A1, one obtains the following remarkable relations in the symmetric approximation,
\be
G^p_M(Q) = \mu_p G^p_E(Q), ~~G^n_M(Q) = \mu_n G^p_M(Q), ~~G^n_E(Q) = 0 .
\label{eq.22}
\ee
In this way one can introduce the standard ff in the symmetric approximation: $G_E^p = F_0(Q)$, which is the basic element for all form factors, so that all others can be expressed via $F_0$, e.g. below with $\tau = Q^2/4M^2$
$$
G_E = F_1 - \tau F_2 = F_0,~~ G_M^p = 2.79 F_0 = F_1^p + F_2^p,~~ G_M^n = -1.91 F_0,$$\be
F_1^p = F_0 \frac{1 + 2.79 \tau}{1 + \tau},~~F_2^p = F_0 \frac{1.79}{1 + \tau},
 F_1^n = -1.91 F_0 \frac{\tau}{1+ \tau},~~ F_2^n = -1.91 \frac{F_0}{1+ \tau}.
 \label{eq.23}
 \ee

 To complete these results we are giving below the form of $F_0(Q)$ in the Breit frame in  general case,
 \be
 F_0 = \sum_i \frac{e_i}{e} \int \frac{d^3 q d^3 k}{(2\pi)^6} \varphi_{(Q/2)} (\veq,\vek)
 \varphi_{(-Q/2)}(\veq +\nu_1^{(i)} \veQ,\vek + \nu_2^{(i)} \veQ)
 \label{eq.24}
 \ee
 where $\varphi_P$ is the baryon wave function with the total momentum $P$. Now with account of the Lorentz contraction  of the boosted wave functions (see (\ref{eq.6})) one obtains the following form of $F_0$, expressed via the rest frame wave functions, namely,
 \be
 F_0(Q) = C_0 \sum_i \frac{e_i}{e} \int \frac{d^2q d^2k d\chi d\kappa}{(2\pi)^6}\varphi(\veq_\bot,\vek_\bot,\kappa,\chi)
 \varphi(\veq_\bot,\vek_\bot,\kappa + \nu_1^{(i)} Q C_0,\chi + \nu_2^{(i)} Q C_0).
\label{eq.25}
\ee
In what follows we shall compare our relations between the ffs, given in (\ref{eq.22}) and (\ref{eq.23}),  which do not depend on the value of $F_0$, with experimental data and as a second step, we shall use the concrete form of nucleon wave functions:
\begin{description}

\item{1)} taken in a simplified Gaussian form with the only parameter for all ff's;

\item{2)} we use the lowest $(K=0)$ hyperspherical wave function for $F_0$.
\end{description}

\section{The symmetric form factors vs experimental data}

The analysis of the baryon wave functions with $L=0$ made in \cite{31,32} shows, that the interaction in
the $\rho$ space is close to an oscillator form, and the resulting wave functions can be well approximated by the Gaussian form in the $\rho$ space and hence also in the its Fourier counterpart $p^2=\veq^2 +
\vek^2$. In principle one can start exact calculation of the baryon wave function $\phi(\veq,\vek)$
using the Hamiltonian as in \cite{31,32} and expanding the wave function as a series in a full set of
orthogonal functions of the Gaussian form, obtaining in this way the exact answer for the wave functions and ff in the given formalism. Having this goal in mind for the future we  below will test the Gaussian form for all nucleon ff with the only parameter in the hypercentral approximation and with the second
numerical parameter in the next $K=2$ approximation.
 To test the resulting ff expressions it is convenient to use the following simple Gaussian form for the wave functions,
\be
\varphi(\veq,\vek) = N \exp \left( - \frac{\veq^2 + \vek^2}{2 \lambda^2} \right), ~~N = (2\sqrt{\pi}/\lambda)^3.
\label{eq.26}
\ee
As a result, the Lorentz contracted nucleon ff has a very simple form,
\be
F_0(Q) = (G_E^p(Q)) = C_0 \exp \left( - \frac{Q^2 M^2 ((\nu_1^{(i)})^2 + (\nu_2^{(I)})^2)}{\lambda^2 (Q^2 + 4M^2)} \right)
\label{eq.27}
\ee
where $C_0 = \sqrt{1-v^2} = \frac{M}{\sqrt{M^2 + Q^2/4}}$. Now using $e_i/e = 2/3,2/3,-1/3$, and $\nu_1^{(i)} = 1/\sqrt{6},1/\sqrt{6},-\sqrt{2/3}$ and
$\nu_2^{(i)} = - 1/\sqrt{2}, 1/\sqrt{2}, 0$, one obtains the final form,
\be
F_0(Q) = C_0 \exp\left( - \frac{2 Q^2 M^2}{3 \lambda^2 (Q^2 + 4 M^2)}\right). \label{eq.28}
\ee

One can easily see that at large $Q$ the ff $F_0 = O(1/Q)$ is different from the perturbative asymptotics \cite{13},
but as we shall see in the region of $Q^2$ below 15 GeV$^2$ it is comparable in the magnitude. It is possible that
this Lorentz type behavior of the form factors is replaced by the perturbative one \cite{13} at some larger $Q^2$,
but below we shall study only the Lorentz contracted form.

One of the first checks of our results, independent of Lorentz contraction, is the ratio $R=\frac{\mu_p G_E^p}{G_M^p}$, which in our symmetric case is equal to unity.
At this point one should compare our predictions with experiment, where an interesting turn in the experimental results, discussed in the Introduction, has taken place. Indeed the experimental data,   using the Rosenbluth type of analysis, could not ensure a good quality of data for the proton electric form factor $G^p_E$, unless another type of experiments with double polarization observables has solved this problem (see e.g. \cite{16} for results and details of experiment). In the non-polarized experiments, based on the analysis of the cross sections, the ratio $\frac{\mu_P G^p_E}
{G^p_M}$ was approximately close to unity, however, with large error bars for $Q^2 \ga 2$ GeV$^2$, which  approximately in agreement with our hypercentral result (\ref{eq.22}). At the same time another relation of (\ref{eq.22})), $\frac{G^p_M}{\mu_p} = \frac{G^n_M}{|\mu_n|}= F_0(Q)$, is valid with accuracy around (10-15)\%, as it can be seen in the data presented in Fig. 18 of \cite{8}. The same situation can be seen in Fig.20 of \cite{8}, which enables us to stress that the
unpolarized experimental data support the hypercentral form factor predictions of (\ref{eq.22}):
all three normalized form factors are roughly equal to each other in unpolarized experiments, while the fourth, $G^n_E$ is zero within 5\% of accuracy. However, in the recoil polarization data of \cite{14,15,16} this ratio R(Q) tends to zero for $Q^2 \cong 8$ GeV$^2$. Indeed this ratio was approximated in \cite{16}
as $R(Q^2)= 1.0587 - 0.14265~ Q^2$. This point and the vanishing of $G^n_E$ in the hypercentral approach requires us to consider the next terms in the hyperspherical expansion (\ref{eq.12}), which is done in the next sections.

\section{The baryon form factors in the hyperspherical basis}

Any wave function of two relative coordinates $\xi,\eta$, defined in (\ref{eq.12}), can be expanded in a series of the hyperspherical functions, suggested and studied  both in the nonrelativistic nuclear physics problems \cite{25,24,26} and in the relativistic baryon problems \cite{27,28,29,30}. In general the expansion has the simple form, shown in the (\ref{eq.12}), and the first two terms can be written as
 \be
 \Psi(\vexi,\veta)=  \psi_0 (\rho) u_0(\Omega) + \psi_2 (\rho) (u'_2(\Omega) + u^{''}_2(\Omega))
 \label{eq.29}
 \ee
 and the hyperspherical ``angular" functions $u'_2, u^{''}_2$ can be written as
 \be
 u'_2= \sqrt{\frac{32}{\pi}} \frac{\vexi \veta}{\rho^2},~~ u^{''}_2=\sqrt{\frac{8}{\pi}} \frac{\vexi^2- \veta^2}{\rho^2}. \label{eq.30} \ee

Here $K= 0,2$ is the power of the $(\frac{\xi}{\rho},\frac{\eta}{\rho})= (\Omega)$ coordinates in $u_K^\nu$, where $\rho=\sqrt{\xi^2 +\eta^2}$ and $\nu$ denotes all other quantum numbers. In the previous section we assumed that it is enough to keep only the first term with $K=0$ and more than that, $\Psi$ can be written in the Gaussian form, similar to the case of the meson form factors \cite{39}.

An important result of the hypercentral ($K=0$) approximation is existence of the unique form factor $F_0(Q)$, defining all three nonzero nucleon
form factors $G^p_E, G^p_M, G^n_M$, as in (\ref{eq.19}), while $G^n_E$ vanishes in this approximation.
In experiment these properties have been well supported by the unpolarized data, while the latest JeffLab experiments \cite{14,15,16} clearly show the decreasing $R(Q)$.
The strong decrease of  $R(Q)$ with $Q$ and nonzero (however, small) values of $G_n^E$ imply that the next term with $K=2$ can be important. Indeed as was mentioned in \cite{17},  in the polarization transfer experiments one must take into account the nonzero internal quark orbital momenta, which are absent in the $K=0$ terms (the hypercental approximation) but present in the $K=2$ case and higher hyperspherical terms.

Now since the $K=0$ term is symmetric in spatial coordinates (or momenta) and does not also contain
any angular momenta, one needs to write the $K=2$ term, which is not symmetric, in the combination with the non-symmetric spin-flavor
parts of the wave function to satisfy the representation (\ref{eq.29}). We can write this representation in the momentum space as follows,
\be
\Psi(q,k;\sigma,{\rm flavor})= \phi_0(q,k)\psi^{\rm(sym)}(\sigma,f) + \phi'_2(q,k)\psi'(\sigma,f) + \phi^{''}_2(q,k)\psi^{''}(\sigma,f).
\label{eq.31} \ee

The spin-flavor functions $\psi(\sigma,f)$ are given in the Appendix A1. Finally, the general form of the nucleon form factor, following (\ref{eq.31}), can be written as follows (keeping only first order terms in $\phi',\phi^{''}$ )
$$
G^N_{E,M}(Q)=\sum_i \int \frac{d^3qd^3k}{(2\pi)^6} \phi_0(q,k)X(p,n|E,M)\phi_0(q+\nu_1^i Q,k+\nu_2^i Q) +$$ $$+
2\sum_i \int \frac{d^3qd^3k}{(2\pi)^6} \phi_0(q,k)[Y'(p,n|E,M)\phi'(q+\nu_1^i Q,k+\nu_2^i Q)+$$\be +
Y^{''}(p,n|E,M)\phi^{''}(q+\nu_1^i Q,k+\nu_2^iQ)]
\label{eq.32}, \ee
Note that in the product of two functions $\phi(q,k)$ in the Breit frame the left wave function  is moving with momentum $-Q/2$, while the right one with momentum $Q/2$.
Here the coefficients $X,Y',Y{''}$ are defined as
$$
X(N,E) = \frac{e_i}{e} \psi^{\rm sym}_N(\sigma,f) \psi^{\rm sym}_N(\sigma,f),~~Y^{''}(N,E) =$$\be =\frac{e_i}{e} \psi^{''}_N(\sigma,f) \psi^{\rm sym}_N(\sigma,f),~~
Y'(N,E) =\frac{e_i}{e}\psi'_N(\sigma,f) \psi^{\rm sym}_N(\sigma,f), N=p,n.
\label{eq.33} \ee
In an analogous way the magnetic matrix elements $X(N,M),$ $Y'(N,M),Y^{''}(N,M)$ are defined, where one must replace in (\ref{eq.33}) $\frac{e_i}{e}$ by $\frac{M e_i}{\omega e} \sigma_i^z$. As a result one obtains the coefficients, shown  below (see explicit forms of all spin-flavor functions in the Appendix A1).

$$ X(p,E) = 1/3, ~Y'(p,E)= Y^{''}(p,E)= -\frac{1}{6\sqrt2}~,~~$$ $$ X(n,E)= 0, ~~Y'(n,E)= Y^{''}(n,E)= \frac{1}{6\sqrt2}~,$$$$
X(n,M) = -\frac{2M}{9\omega},~~ Y'(n,M)= -\frac{7M}{18\omega\sqrt6}~,~~ $$ $$Y^{''}(n,M)= -\frac{M\sqrt2}{12\omega}~,~
X(p,M)=\frac{M}{3\omega}~,~~$$\be Y'(p,M)=\frac{M}{\omega 6\sqrt6}~,~~ Y^{''}(p,M)= \frac{M}{\omega 18\sqrt2}.
\label{eq.34},
\ee
Thus we have defined all four nucleon form factors in terms of their known hyperspherical wave functions
with $K=0,2$, where we take into account the $K=2$ terms in the lowest (first) order. The resulting form and comparison with data is the topic of the next section.

\section{The hyperspherical form factor representation vs data. Comparison with other approaches}

Let us start with the definition of the basic form factor elements in  the hyperspherical basis.
For $K=0$ one has the term in (\ref{eq.25}) which contains only $K=0$ wave functions (the hypercentral approximation),
\be
f_0(Q)= \sum_i\int \frac{d^3q d^3k}{(2\pi)^6}\phi_{-Q/2}^{(0)}(q,k)\phi_{Q/2}^{(0)}(q+\nu_1^i Q,k+\nu_2^i Q).
\label{eq.35} \ee
One can also define the functions $\phi'(q,k),\phi^{''}(q,k)$ as follows
\be
\phi'(q,k)= \sqrt\frac{32}{\pi}\veq\vek \phi^{(2)}(q^2+k^2), ~~\phi^{''}(q,k)= \sqrt\frac{8}{\pi}(q^2-k^2)\phi^{(2)}(q^2+k^2).
\label{eq.36} \ee
At this point one can define the $K=2$ admixture in the form factors as follows
\be
f'_2(Q)= \sum_i \int\frac{d^3q d^3k}{(2\pi)^6}\phi_{-Q/2}^{(0)}(q,k)\phi'_{Q/2}(q+\nu_1^i Q,k+\nu_2^i Q)
\label{eq.37} \ee
and similar definition of $f^{''}_2(Q)$, where instead of $\phi'$ one inserts $\phi^{''}$.

To take into account the electric and the magnetic charge matrix elements between spin-flavor wave functions we introduce
the coefficients $a'_i,a^{''}_i,b'_i,b^{''}_i,$ $c'_i,c^{''}_i,d'_i,d^{''}_i$ for the matrix elements between $\psi^{\rm sym}(\sigma,f)$ and
$\psi',\psi^{''}$ for $G^n_E,G^p_E,G^n_M,G^p_M$, respectively, e.g.
\be
(\psi^{\rm sym}_n(\sigma,f)\frac{e_i}{e}\psi'_n(\sigma,f)= (a'_1,a'_2,a'_3)=\frac{1}{2\sqrt6},
 -\frac{1}{2\sqrt6},0,
\label{eq.38} \ee
and similar definitions for $b,c,d$, see Appendix A1.
As a result one obtains
$$
G^n_E(Q)=2 \sum_i \int \frac{d^3qd^3k}{(2\pi)^6}\phi_{-Q/2}^{(0)}(q,k)(a'_i\phi'_{Q/2}(q+\nu^i_1 Q,k+\nu^i_2 Q) + $$\be + a^{''}_i\phi^{''}_{Q/2}(q+\nu^i_1 Q,k+\nu^i_2 Q)), \label{eq.39} \ee
$$
G^p_E(Q)=\sum_i \int \frac{d^3qd^3k}{(2\pi)^6}\phi_{-Q/2}^{(0)}(q,k)$$
\be \left[1/3 \phi_{Q/2}^{(0)}(q+\nu^i_1 Q,k+\nu^i_2 Q) -
2(b'_i\phi'_{Q/2}(q+\nu^i_1 Q,k+\nu^i_2 Q) + b^{''}_i\phi^{''}_{Q/2}(q+\nu^i_1 Q,k+\nu^i_2 Q))\right], \label{eq.40} \ee
$$
G^n_M(Q)= \frac{M}{\omega} \sum_i \int \frac{d^3qd^3k}{(2\pi)^6}\phi_{-Q/2}^{(0)}(q,k)$$\be\left[\left(-\frac{2}{3}\right)\phi_{Q/2}^{(0)}(q+
\nu^i_1 Q,k+\nu^i_2 Q)+ 2(c'_i\phi'(q+\nu^i_1 Q,k+\nu^i_2 Q)+ c^{''}_i\phi^{''}(q+\nu^i_1 Q,k+\nu^i_2 Q)\right],
\label{eq.41} \ee
$$
G^p_M(Q)=M/\omega \sum_i \int \frac{d^3qd^3k}{(2\pi)^6}\phi_{-Q/2}^{(0)}(q,k)$$\be\left[\phi_{Q/2}^{(0)}(q+\nu^i_1 Q,k+\nu^i_2 Q)+2(d'_i\phi'(q+\nu^i_1 Q,k+\nu^i_2 Q)+d^{''}_i\phi^{''}(q+\nu^i_1,k+\nu^i_2 Q)\right]. \label{eq.42} \ee
The numerical values of the coefficients $a'_i, b'_i, ...$ are given in the Appendix A1.

We now turn to exact form of the $K=2$ wave functions, defined in (\ref{eq.36}), and first  we shall  assume a simple Gaussian form for both normalized $\phi^{(0)}, \phi^{(2)}$ with the only parameter $\lambda$, as in (\ref{eq.27}), so that the functions $\phi'(q,k), \phi^{''}(q,k)$ contain an admixture parameter $A$, measuring the contribution (amplitude) of the $K=2$ functions in the total nucleon wave function. Then the assumed trial wave functions $\phi', \phi^{''}$ have the form
\be
\phi'(q,k)= N' \veq\vek \exp\left(-\frac{q^2 +k^2}{2\lambda^2}\right), ~~\phi^{''}(q,k)= N^{''} (q^2 -k^2) \exp\left(-\frac{q^2+k^2}{2\lambda^2}\right),~\label{eq.43} \ee
where $N'= A \frac{32\pi\sqrt2}{\lambda^3}$  and $N^{''}= A \frac{8\pi^2}{\lambda^3 \sqrt{3\pi}}$.
As a result the $K=2$ contribution to the form factors acquires the form
$\Delta F(Q^2)= \frac{2A Q^2}{\lambda^2 4\sqrt{3}} K^{E,M}_{p,n} \exp(-\frac{Q^2}{6\lambda^2})$,
where we have taken into account that $(\nu^i_1)^2 + (\nu^i_2)^2= \frac{2}{3}$ for every i, and the coefficients $K$ defined as, e.g., $K^M_p= \sum_i(d'_1 2\nu^i_1\nu^i_2 + d^{''}_2 ((\nu^i_1)^2 - (\nu^i_2)^2))$
\be
K^M_n= \frac{\sqrt{ 2}}{3},~ K^M_p= -\frac{\sqrt 2}{9},~ K^E_n=-\frac{2}{3\sqrt 2},~ K^E_p=\frac{2}{3\sqrt 2}.
   \label{eq.44} \ee

Then in this simple approximation with two parameters $A,\lambda$ the nucleon form factors are
$$
G^p_E= \exp\left(-\frac{Q^2}{6\lambda^2}\right) \left(1 + A\sqrt{3}\frac{Q^2}{6\lambda^2} K^E_p\right),
G^n_E= \exp\left(-\frac{Q^2}{6\lambda^2}\right) A\sqrt{3} \frac{Q^2}{6\lambda^2} K^E_n,$$\be
G^p_M= \mu_p \exp\left(-\frac{Q^2}{6\lambda^2}\right)\left(1 + A\sqrt{3} \frac{Q^2}{6\lambda^2} K^M_p\right),
G^n_M= \mu_n \exp\left(-\frac{Q^2}{6\lambda^2}\right)\left(1 - A\sqrt{3} \frac{3Q^2}{12\lambda^2} K^M_n\right).
\label{eq.45} \ee

Here the proton and neutron magnetic moments are taken at their physical values.
As it is seen in (\ref{eq.45}) the coefficient $A$ can be found from the ratio $R(Q)$, which was found experimentally in \cite{16}
\be
R(Q)= 1.0587 - 0.14265 Q^2.
\label{eq.46} \ee

This defines the ratio $\frac{A}{6\lambda^2}= - 0.27$ and now all form factors are expressed via  one parameter $\lambda$ and can be written as,
$$
G^p_E= f(Q)(1-0.11 Q^2), ~~\frac{G^M_p}{\mu_p}= f(Q)(1+0.0367 Q^2), ~~$$\be\frac{G^n_M}{\mu_n}= f(Q)(1+0.165 Q^2), ~~G^n_E= f(Q) 0.11 Q^2, ~~f(Q)=\exp\left(-\frac{Q^2}{6\lambda^2}\right). \label{eq.47} \ee
One can compare these expressions with experimental data and persuade himself that indeed  the contribution of the terms with $K=2$ has correct qualitative features: $G^p_E < G^p_M/\mu_p$ for negative $K=2$ amplitude $A$, whereas $G^n_E > 0$ and small for the same $A$. Also $G^n_M$ and $G^p_M$ are close to each other as in data, however, the $Q^2$ dependence of $f(Q)$, as given in (\ref{eq.47}), contradicts data for any value of $\lambda$. At this point we turn back to our discussion of the Lorentz contracted baryon wave functions and persuade ourselves that the ff with the Lorentz contracted wave functions should depend on $Q^2$ as $C_0 F(C_0^2 Q^2)$. Correspondingly we modify our expressions in (\ref{eq.45}), multiplying all $Q^2$ factors by the contraction factor
$C_0^2= \frac{4M^2}{4M^2 + Q^2}$ and moreover, multiplying the full function $f(Q)$ by $C_0$, as it is
prescribed in (\ref{eq.20}), (\ref{eq.28}). The result can be written in the following way,
\be
G^p_E= \bar f(Q) \left(1 + \sqrt{\frac{2}{3}} \frac{A}{6\lambda^2} g(Q^2)\right), \label{eq.48} \ee

\be
\frac{G^p_M}{\mu_p}= \bar f(Q)\left (1 -\frac{1}{3}\sqrt{\frac{2}{3}} \frac{A}{6\lambda^2} g(Q^2)\right), \label{eq.49} \ee

\be G^n_E= -\sqrt{\frac{2}{3}} \frac{A}{6\lambda^2} \bar f(Q) g(Q^2), \label{eq.50} \ee

\be
\frac{G^n_M}{|\mu_n|}= \bar f(Q) (1 - \sqrt{\frac{3}{2}}\frac{A}{6\lambda^2} g(Q^2), \label{eq.51} \ee
where the following notations are  used,
\be
\bar f(Q)= C_0 \exp(-\frac{2M^2 Q^2}{3\lambda^2(Q^2+4M^2)}),~~ g(Q^2)= \frac{Q^2 4M^2}{Q^2+4M^2}. \label{eq.(52)} \ee
One can now compare the ratio $R(Q^2)$ from the ratio of  (\ref{eq.47}) and (\ref{eq.48}) with the
experimental data, given in \cite{16}, Table (4), to define the ratio $\frac{A}{6\lambda^2}$, which yields
\be
\frac{A}{6\lambda^2}= -(0.27 \pm 0.025)~~ {\rm GeV}^{-2}, \label{eq.(53)} \ee
One can see the resulting ratio $R^{\rm th}(Q)$ with the experimental data on $R(Q)$ from \cite{16} given in the Table~\ref{tab.01}. Note that the data of \cite{16} are obtained in the polarization transfer
method,which allows to obtain accurate values of $G^p_E$ in contrast to the Rosenbluth method.


\begin{table}[!htb]
\caption{The ratio $\frac{\mu_p G^p_E}{G^p_M}$ as a function of $Q^2$ from \cite{16} in comparison with the $K=0;2$  hyperspherical approximation}
\begin{center}
\label{tab.01} \begin {tabular}{ |c|c|c|c|c|}\hline
$Q^2$(GeV$^2)$ & 3.5 & 4.0 & 4.8 & 5.6 \\\hline
$R(Q)$ & $0.571\pm $&$ 0.517\pm $ &$ 0.450 \pm $ &$ 0.354 \pm $ \\
&$
0.072 \pm 0.007$ &$  0.056\pm 0.009$ &$ 0.05 \pm 0.012$ &$0.055\pm 0.019$ \\\hline
$R^{(th)}(Q)$ & 0.541 & 0.516 & 0.481 & 0.45 \\\hline
\end{tabular}
\end{center}
\end{table}
One can see that the higher $Q^2$ values prefer a somewhat larger value $\frac{A}{6\lambda^2}= -0.33$,
which we shall take into account reproducing the $G_E^n$ form factor.
Finally, to define the parameter $\lambda$ one can compare $R_M(Q)$=$ \frac{G^p_M}{\mu_p G_D}$ with $G_D$=$\frac{1}{(1 + \frac{Q^2}{0.71 GeV^2})^2}$ and $G^p_M(Q)$ from (\ref{eq.49}) with the data from \cite{46} which is given in Table~\ref{tab.02} below.


\begin{table}[!htb]
\caption{ The ratio $R_M(Q)$ = $\frac{G^p_M}{\mu_p G_D}$ as a function of $Q^2$ from \cite{46} in comparison with the $K=0;2$ hyperspherical approximation}
\begin{center}
\label{tab.02} \begin{tabular} { |c|c|c|c|c|c|} \hline
$Q^2$(GeV$^2$) & 2.2 & 2.75 & 3.75 &4.20 & 5.20 \\\hline

$R_M(Q)$ from \cite{46} &$ 1.05 \pm$ & $1.055 \pm $ & $1.044 \pm$ & $1.012 \pm$ & $1.007 \pm $ \\

 &$  0.016$ & $ 0.018$ & $ 0.015$ & $ 0.012$ & $0.032$ \\\hline

$R^{(th)}_M(Q)$ & 1.12 & 1.07 & 0.98 & 0.997 & 0.975 \\\hline
\end{tabular}
\end{center}
\end{table}

It is interesting also to compare the resulting ff $G^p_M(Q)$ with the data at larger values of $Q^2$
which have been measured recently in the Jefferson Lab (Hall A)  \cite{51} by the Rosenbluth method,
which is sensitive to the magnetic ff. These results are given below in Table 3.

\begin{table}[!htb]
\caption{ The ratio $R_M(Q)$ = $\frac{G^p_M}{\mu_p G_D}$ as a function of $Q^2$ from \cite{47} in
comparison with the $K=0;2$ hyperspherical approximation}
\begin{center}
\label{tab.03} \begin{tabular} { |c|c|c|c|c|c|c|} \hline
$Q^2$(GeV$^2$) & 5.99 & 7.02 & 7.94 & 8.994 & 9.84 & 12.25 \\\hline

$R_M(Q)$ from \cite{47} &$ 1 \pm$ & $ 0.967 \pm $ & $0.943 \pm $ & $0.934 \pm $ & $0.909 \pm$ &$ 0.858 \pm $ \\
&$ 0.011$ & $ 0.015 $ & $ 0.018 $ & $ 0.016 $ & $ 0.029 $ & $ 0.019 $ \\\hline

$R^{(th)}_M(Q)$ & 0.979 & 0.977 & 0.969 & 1.017 & 1.0228 & 1.094 \\\hline
\end{tabular}
\end{center}
\end{table}

Using these data one can find the coefficient $\frac{1}{6\lambda^2}= 1.88 $ GeV$^{-2}$, which yields using (\ref{eq.(53)}) $A= -0.143$. In this way we have defined both numerical constants entering in our four
nucleon ff (\ref{eq.48}-\ref{eq.(52)}), and we can now check the results for $G^n_M(Q)$ and $G^n_E(Q)$.
We start with  the electric neutron form factor $G^n_E$, which is zero in the hypercentral ($K=0$) approximation and is directly proportional to the admixture of the $K=2$ amplitude $A$,
\be
G^n_E(Q)= - \sqrt{\frac{2}{3}} \frac{A}{6\lambda^2} g(Q^2) \bar f(Q) . \label{eq.54)} \ee

Introducing here the known values of the factors one obtains for $Q^2= 1.45$ GeV$^2$ the estimate
$G^n_E(1.45$ GeV$^2)= 0.027$, which can be compared to experimental number from \cite{48}, equal to
$ 0.038 \pm 0.011$ in a reasonably good agreement. In a similar way the experimental data from \cite{49}
for $Q^2$=$0.59 GeV^2$ yields $G^n_E$=$ 0.048 \pm 0.0052$, while our $G^n_E(Q)$ yields $0.0397-0.0479$ (the larger value is for larger $A= -0.175$). At small $Q^2$ we obtain the behavior  $G^n_E(Q^2)= 0.022 Q^2 + O(Q^4)$ which also is in agreement with data. Note, that the experiment \cite{48} was done in the polarization transfer method, which in this example is again connected with the $K=2$ contribution.

We can now consider the case of the ff $G^n_M$ and again to stress the possible difference of results in the standard cross section (the Rosenbluth) and the polarization transfer methods, the latter have been used mostly for low $Q < 1$ GeV \cite {50}. In the region $Q^2=(1-5)$~GeV$^2$ the values of  $R^n_M= \frac{G^n_M}{\mu_n G_D}$, measured in experiment, occur in the interval $R^n_M= (0.9 - 1.1)$. This can  be compared with our result in the $K=0$ (hypercentral) approximation, $R(Q)= \frac{\bar f(Q)}{G_D}$, yielding in this region a slow decrease from the number 1.2 to 0.9 . At $Q^2=8$ GeV$^2$  data from \cite{51} give $R(8$ GeV$^2)= 0.7 \pm 0.11$, while our result without the orbital excitations $K=2$ components yields $R(Q)= 0.79-0.83$ in the range  $Q^2= (5.6- 8.0)$ GeV$^2$. As one can see, the agreement is reasonable, taking into account that we have only two fixed parameters $A,\lambda$ to describe four nucleon form factors  in the whole region of $Q^2$ from zero to 8-10 GeV$^2$. Especially interesting is the behavior of the basic ratio $\bar R(Q)= \frac{\bar f(Q)}{G_D}$, which drops beyond $Q=5$ GeV$^2$ in agreement with data for both $G^p_M$ and $G^n_M$ and is close to unity for smaller values of $Q$. As it is, the admixture of the $K=2$ component is crucial for the decrease of $G^p_E(Q)$ and gives nonzero positive values of $G^n_E$, being zero in the $K=0$ approximation.
Summarizing, the hypercentral theory of nucleon ff without any parameters in a general form ensures the equality
of 3 nucleon normalized ff (except $G^n_E$) in the low $Q$ region which agrees with data within 10\% of accuracy, while $G^n_E$ is zero in the same range of accuracy. Now adding $K=2$ terms one has a possibility of a positive nonzero $G^n_E$ and at the same time of the decreasing ratio $\frac{G^p_E}{G^p_M}$,while the ratio $\frac{G^n_M}{G^p_M}$ remains of the order of unity. Introducing two real parameters $A,\lambda$
and choosing the Gaussian form one arrives at the hyperspherical nucleon ff which agree with data with accuracy around 10 \% for $Q^2 < 8 GeV^2$. It was not a purpose of the paper to provide a few percent
accurate many-parameter reproduction of the experimental data, like it is done in the well known Kelly form \cite{51*} with around 20 parameters, but the hyperspherical approach is aimed at the background understanding of the dynamics of the nucleon ff.

\section{Conclusions and an outlook}

There are two main points of the paper -- the validity of the Lorentz contraction phenomenon  and the role of higher hyperspherical components in the nucleon ff which are presented in six items below.

\begin{description}

\item{1)} The three-body structure of baryons is treated correctly in the hyperspherical formalism, which is especially convenient because of the three-body character of the main forces in the system (each quark is connected by the string to the string junction point), which are  not of two-body character. As a result one has a good prediction of the basic hypercentral form factor $f_0(Q)$ such that the relation  $\frac{G^p_M}{\mu_p}= \frac{G^n_M}{\mu_n}= G^p_E = f_0(Q)$ is fulfilled approximately at  $Q^2$ below $2~GeV^2$.

\item{2)} The basic discovery of the polarization transfer experiments -- the strongly decreasing ratio $R(Q)= \frac{\mu_p G^p_E}{G^p_M}$ can be explained by the next terms with $K=2$ of the hyperspherical formalism. These terms contain nonzero orbital components $l_\xi= l_\eta =1,~L= 0$ and both the sign and the magnitude of the constant $A$ of this correction, can be defined from $R(Q)$.

\item{3)} Another basic result, obtained in the hyperspherical formalism, is the vanishing of the neutron charge form factor $G^n_E$ in the leading hypercentral approximation $K=0$  and its proportionality to the $K=2$ admixture amplitude $A$, so that $G^n_E$ can be presented as $(-A)$ times positive function. It was checked above in the paper that this property and the resulting value of $G^n_E$ agrees with experiment. Moreover, the behavior $G^n_E(Q)= {\rm const}~ Q^2$ for small $Q$ predicted by the method also agrees with experimental data.

\item{4)} The explicit $Q^2$ dependence of the basic form factor $f_0(Q)$ and its $K=2$ correction is not predicted exactly from the theory in the paper, but it is assumed for simplicity in the Gaussian form with the only free parameter $\lambda$, while the $K=2$ correction proportional to $Q^2$ contains another free parameter $A$. The relativistic hyperspherical formalism described above using Hamiltonian formalism of \cite{30,31,32} allows to obtain both wave functions and ff of the
    nucleon and this serious goal is planned for the future,where it will be possible to compare our results both with experiment and with other existing theoretical methods, e.g. \cite{11,12,52}.
\item{5)} Another basic goal of this paper is the analysis of the role of the Lorentz contraction mechanism for the strongly accelerated baryon wave functions, which should be present in the
    instant time formalism of QCD, as it was explained in section 2. The strongly contracted wave function is insensitive to the acceptance of large momentum $Q$ and this fact determines the asymptotics of the nucleon ff. As it was shown in the previous section this behavior is roughly in agreement with the data for $G^p_M(Q)$ up to $Q^2=15 $ GeV$^2$. This topic discussed previously in
\cite{36,37,38,39,40} has shown the reality and the importance of this contraction effect,and the present paper gives additional information.
\item{6)} The connection of the Rosenbluth cross section  data and the polarization transfer data to
the hypercentral and the higher K-harmonics approximation respectively  was mentioned but not discussed properly in the paper. The physical reason for the emergence of the $K=2$ term which
provides a needed decrease of $G^p_E(Q)$ is the appearance of two $l=1$ angular momenta in the $3q$
system, which sum up into $L=0$ total angular momentum. It is argued in \cite{17} that the helicity flip processes in the polarization transfer experiments are closely connected with the nonzero quark
orbital momenta and in this way the $K=2$ component can be basically important in these experiments.
Another feature of the $3q$ baryon dynamics to explain the polarization transfer experiments suggested in \cite{17*} is the effective repulsive core which can be also connected with the nonzero quark angular momenta, which effectively decrease the wave function at the origin.
The elucidation of this very interesting topic  requires a  detailed
 analysis of the spin-momentum structure of polarization and cross section data and will be delayed for future publications.

\end{description}
The author is grateful for help and useful discussions  to A.M.Badalian.

This work was supported by the Russian Science Foundation in the framework of the scientific project,
grant 16-12-10414.

\newpage
\vspace{1cm}

\vspace{1cm}

{\bf Appendix A1.}

 {\bf The spin-flavor wave functions for proton and neutron}

 \setcounter{equation}{0} \def\theequation{A1.\arabic{equation}}

Below we denote the $u,d$ quarks with spins up or down as $u_+,u_-,d_+,d_-$.
Neutron.
$$
\Psi^{\rm(sym)}_n (\sigma,f)= \frac{1}{2\sqrt3} ( 2u_-d_+d_+  - d_-u_+d_+  - u_+d_-d_++ $$
\be + 2 d_+u_-d_+  -d_+d_-u_+
-d_-d_+u_+  -d_+u_+d_-  -u_+d_+d_-  +2 d_+d_+u_-),\label{eq.A1.1}\ee

$$
\Psi'_n(\sigma,f)= \frac{1}{2\sqrt3} ( d_-d_+u_+  -d_+d_-u_+  -2 d_+d_-u_+ +$$ \be +2 u_-d_+d_+   +u_+d_+d_- -d_+u_+d_-)\label{eq.A1.2}\ee

 $$
\Psi^{''}_n(\sigma,f)= \frac{1}{6} (-d_+d_-u_+  -d_-d_+u_+  -4 d_+d_+u_-  +2u_+d_-d_+ +$$\be +2d_-u_+d_+  -d_+u_+d_-
+2 U_-d_+d_+  -u_+d_+d_-   +2 u_+d_-d_+ )\label{eq.A1.3}\ee

Proton
$$
\Psi^{\rm(sym)}_p (\sigma,f)= \frac{1}{3\sqrt2} ( 2 u_+d_-u_+  + 2 d_-u_+u_+  -d_+u_-u_+  -u_-d_+u_+   -u_+u_-d_+
-u_-u_+d_+ -$$\be  -u_+d_+u_-  -d_+u_+u_-   +2 u_+u_+d_-),\label{eq.A1.4}\ee

\be
\Psi'_p(\sigma,f)= \frac{1}{2\sqrt3} ( u_-u_+d_+   -u_+u_-d_+   -2 u_+d_-u_+  +2 d_-u_+u_+   + d_+u_+u_-
 - u_+d_+u_- ),\label{eq.A1.5}\ee

 \be
\Psi^{''}_p (\sigma,f)= \frac{1}{6} ( - u_+u_-d_+  -u_-u_+d_+   -4 u_+u_+d_-  +2 U_+d_-u_+   +2 u_-d_+u_+
-u_+d_+u_- +$$ $$  +2d_-u_+u_+   -d_+u_+u_-  +2 d_+u_-u_+ ).\label{eq.A1.6}\ee

The matrix elements,
\be(a'_i,a^{''}_i,b'_i,b^{''}_i,c'_i,c^{''}_i,d'_i,d^{''}_i)\label{eq.A1.7}\ee

general structure \be a'_i,b'_i,c'_i,d'_i = x_(a,b,c,d) ( 1, -1, 0 ), a^{''}_i,b^{''}_i,c^{''}_i,d^{''}_i= y_(a,b,c,d) (1,1,-2),\label{eq.A1.8}\ee
where

\be
x_(a,b,c,d)= \frac{1}{2\sqrt6}, -\frac{1}{2\sqrt6}, - \frac{1}{2\sqrt6}, \frac{1}{6\sqrt6},\label{eq.A1.9}\ee
and \be
y_(a,b,c,d)= \frac{1}{6\sqrt2}, -\frac{1}{6\sqrt2}, - \frac{\sqrt2}{12}, \frac{1}{18\sqrt2}.\label{eq.A1.10}\ee

\Large

\end{document}